\input{aipcheck.tex}

\documentclass{aipproc}
\usepackage{amssymb,amsmath}

\def\selectedlayoutstyle{6x9}
\layoutstyle\selectedlayoutstyle

\SetInternalRegister\hbadness{8000} 

\newcommand\doingARLO[2][]{%
  \ifx\mmref\undefined #1\else #2\fi
}

\begin{document}

\title
      [Deriving Laws from Ordering Relations]
      {Deriving Laws from Ordering Relations}


\author{Kevin H. Knuth}
{address={Computational Sci. Div., NASA Ames Research Ctr., M/S
269-3, Moffett Field CA 94035}}

\copyrightyear{2003}

\begin{abstract}
The effect of Richard T. Cox's contribution to probability theory
was to generalize Boolean implication among logical statements to
degrees of implication, which are manipulated using rules derived
from consistency with Boolean algebra. These rules are known as
the sum rule, the product rule and Bayes' Theorem, and the measure
resulting from this generalization is probability. In this paper,
I will describe how Cox's technique can be further generalized to
include other algebras and hence other problems in science and
mathematics. The result is a methodology that can be used to
generalize an algebra to a calculus by relying on consistency with
order theory to derive the laws of the calculus. My goals are to
clear up the mysteries as to why the same basic structure found in
probability theory appears in other contexts, to better understand
the foundations of probability theory, and to extend these ideas
to other areas by developing new mathematics and new physics. The
relevance of this methodology will be demonstrated using examples
from probability theory, number theory, geometry, information
theory, and quantum mechanics.
\end{abstract}

\date{\today}
\maketitle

\section{Introduction}

The algebra of logical statements is well-known and is called
\emph{Boolean algebra} \cite{Boole:1848, Boole:1854}. There are
three operations in this algebra: conjunction $\wedge$,
disjunction $\vee$, and complementation $\sim$. In terms of the
English language, the logical operation of conjunction is
implemented by the grammatical conjunction `\emph{and}', the
logical operation of disjunction is implemented by the grammatical
conjunction `\emph{or}', and the logical complement is denoted by
the adverb `\emph{not}'. Implication among assertions is defined
so that a logical statement $a$ implies a logical statement $b$,
written $a \rightarrow b$, when $a \vee b = b$ or equivalently
when $a \wedge b = a$.  These are the basic ideas behind Boolean
logic.

The effect of Richard T. Cox's contribution to probability theory
\cite{Cox:1946, Cox:1961} was to generalize Boolean implication
among logical statements to degrees of implication represented by
real numbers. These real numbers, which represent the degree to
which we believe one logical statement implies another logical
statement, are now recognized to be equivalent to probabilities.
Cox's methodology centered on deriving the rules to manipulate
these numbers. The key idea is that these rules must maintain
consistency with the underlying Boolean algebra. Cox showed that
the product rule derives from associativity of the conjunction,
and that the sum rule derives from the properties of the
complement. Commutativity of the logical conjunction leads to the
celebrated Bayes' Theorem. This set of rules for manipulating
these real numbers is not one of set of many possible rules; it is
the \emph{unique generalization consistent with the properties of
the Boolean algebraic structure}.

Boole recognized that the algebra of logical statements was the
same algebra that described sets \cite{Boole:1848}. The basic idea
is that we can exchange `set' for `logical statement', `set
intersection' for `logical conjunction', `set union' for `logical
disjunction', `set complementation' for `logical complementation',
and `is a subset of' for `implies' and you have the same algebra.
We exchanged quite a few things above and its useful to break them
down further.  We exchanged the \emph{objects} we are studying:
`sets' for `logical statements'.  Then we exchanged the
\emph{operations} we can use to combine them, such as `set
intersection' for `logical conjunction'.  Finally, we exchanged
the means by which we \emph{order} these objects: `is a subset of'
for `implies'.  The obvious implication of this is that Cox's
results hold equally well for defining measures on sets.  That is
we can assign real numbers to describe the degree to which one set
is a subset of another set. The algebra allows us to have a sum
rule, a product rule, and a Bayes' Theorem analog---just like in
probability theory!

It has been recognized for some time that there exist
relationships between other theories and probability theory, but
the underlying reasons for these relationships have not been well
understood. The most obvious example is quantum mechanics, which
has much in common with probability theory. However, it is clearly
not probability theory since quantum amplitudes are complex
numbers rather than real numbers. Another example is the analogy
recognized by Carlos Rodr\'{i}guez between the cross-ratio in
projective geometry and Bayes' Theorem \cite{Carlos:1991}. In this
paper, I will describe how Cox's technique can be further
generalized to include other algebras and hence other problems in
science and mathematics. I hope to clear up some mysteries as to
why the same basic structure found in probability theory appears
in other contexts. I expect that these ideas can be taken much
further, and my hope is that they will eventually lead to new
mathematics and new physics. Last, scattered throughout this paper
are many observations and connections that I hope will help lead
us to different ways of thinking about these topics.

\section{Lattices and Algebras}
\subsection{Basic Ideas}
The first step is to generalize the basic idea behind the elements
described in our Boolean algebra example: objects, operations, and
ordering relations. We start with a set of objects, and we select
a way to compare two objects by deciding whether one object is
`greater than' another object. This means of comparison allows us
to order the objects. A set of elements together with a
\emph{binary ordering relation} is called a \emph{partially
ordered set}, or a \emph{poset}. It is called a \textit{partial}
order to allow for the possibility that some elements in the set
cannot be directly compared. In our example with logical
implication, the ordering relation was `implies', so that if `$a$
implies $b$', written $a \rightarrow b$, then $b$ is in some sense
`greater than' $a$, or equivalently, $a$ is in some sense
`included in' $b$. An ordering relation is generally written as $a
\leq b$, and read as `$b$ includes $a$' or `$a$ is contained in
$b$'. When $a \leq b$, but $a \neq b$ then we write $a < b$, and
read it as `$a$ is properly contained in $b$'. Last, if $a < b$,
but there does not exist any element $x$ in the set $P$ such that
$a < x < b$, then we say that `$b$ covers $a$', and write $a \prec
b$. In this case, $b$ is an immediate successor to $a$ in the
hierarchy imposed by the ordering relation.  It should be stressed
that this notation is general for all posets, but that the
relation $\leq$ means different things in different examples.

\begin{figure}
\label{fig:posets} \caption{Four different posets. (a) The
positive integers ordered by `is less than or equal to'. (b) The
positive integers ordered by `divides'. (c) The powerset of $\{a,
b, c\}$ ordered by `is a subset of'. (d) Three mutually exclusive
logical statements $a$, $k$, $n$ ordered by `implies'. Note that
the same set can lead to different posets under different ordering
relations (a and b), and that different sets under different
ordering relations can lead to isomorphic posets (c and d).}
\includegraphics[height=.45\textheight]{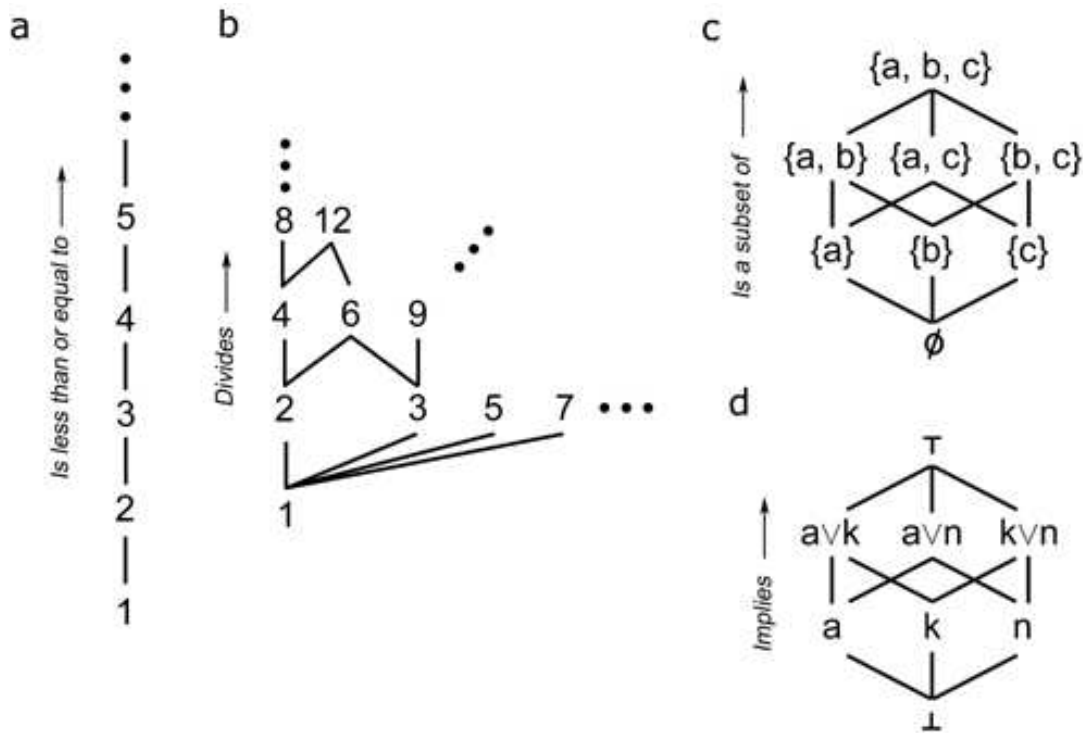}
\end{figure}

We can draw diagrams of posets using the ordering relation and the
concept of covering. If for two elements $a$ and $b$, $a \leq b$
then we draw $b$ above $a$ in the diagram. If $a \prec b$ then we
connect the elements with a line. Figure \ref{fig:posets} shows
examples of four simple posets. Figure \ref{fig:posets}a is the
set of positive integers ordered according the the usual relation
`is less than or equal to'. From the diagram one can see that $2
\leq 3$, $2 \prec 3$, but that $2 \nprec 4$. Figure
\ref{fig:posets}b shows the set of positive integers ordered
according to the relation `divides'. In this poset $2 \leq 8$
means that 2 divides 8. Also, $4 \prec 12$ since 4 divides 12, but
there does not exist any positive integer $x$, where $x \neq 4$
and $x \neq 12$, such that 4 divides $x$ and $x$ divides 12. This
poset is clearly important in number theory. Figure
\ref{fig:posets}c shows the set of all subsets (the
\emph{powerset}) of $\{a, b, c\}$ ordered by the usual relation
`is a subset of', so that in this case $\leq$ represents
$\subseteq$. From the diagram one can see that $\{a\} \subseteq
\{a,b,c\}$, and that $\{b\}$ is covered by both $\{a,b\}$ and
$\{b,c\}$. There are elements such as $\{a\}$ and $\{b\}$ where
$\{a\} \nsubseteq \{b\}$ and $\{b\} \nsubseteq \{a\}$. In other
words, the two elements are incomparable with respect to the
ordering relation. Last, Figure \ref{fig:posets}d shows the set of
three mutually exclusive assertions $a, k,$ and $n$ ordered by
logical implication. These assertions, discussed in greater detail
in \cite{Knuth:PhilTrans}, represent the possible suspects
implicated in the theft of the tarts made by the Queen of Hearts
in Chapters XI and XII of Lewis Carroll's \textit{Alice's
Adventures in Wonderland}
\begin{align*}
a &= \textit{`Alice stole the tarts!'}\\
k &= \textit{`The Knave of Hearts stole the tarts!'}\\
n &= \textit{`No one stole the tarts!'}
\end{align*}
Logical disjunctions of these mutually exclusive assertions appear
higher in the poset. The element $\bot$ represents the absurdity,
and $\top$ represents the disjunction of all three assertions,
$\top = a \vee k \vee n$, which is the truism.

There are two obvious ways to combine elements to arrive at new
elements. The \emph{join} of two elements $a$ and $b$, written $a
\vee b$, is their least upper bound, which is found by finding
both $a$ and $b$ in the poset and following the lines upward until
they intersect at a common element. Dually, the \emph{meet} of $a$
and $b$, written $a \wedge b$, is their greatest lower bound. In
Figure \ref{fig:posets}c, $\{a\} \vee \{b\} = \{a,b\}$, and
$\{a,b\} \wedge \{b,c\} = \{b\}$. In that poset, the join $\vee$
is found by the set union $\cup$ and the meet $\wedge$ by set
intersection $\cap$. For the poset of logical statements in Figure
\ref{fig:posets}d the notation is a bit more transparent as the
join is the logical disjunction (OR). For the meet we have, $(a
\vee k) \wedge (k \vee n) = k$, which is the logical conjunction
(AND). In Figure \ref{fig:posets}b, $\wedge$ represents the
greatest common divisor $gcd()$; whereas $\vee$ represents the
least common multiple $lcm()$. In Figure \ref{fig:posets}a, $2
\vee 3 = 3$, and $2 \wedge 3 = 2$. In this case $\vee$ acts as the
$max()$ function selecting the greatest element; whereas $\wedge$
acts as the $min()$ function selecting the least element. Again,
it is important to keep in mind that the symbols $\vee$ and
$\wedge$ mean different things in different posets.

Some posets possess a top element, which is called \emph{top} and
is generally written as $\top$. The bottom element, called
\emph{bottom}, is generally written as $\bot$ or equivalently
$\varnothing$.  There is an important set of elements in the poset
called the \emph{join-irreducible elements}. These are elements
that cannot be written as the join of two other elements in the
poset. The bottom is never included in the join-irreducible set.
In Figure \ref{fig:posets}c, there are three join-irreducible
elements $\{a\}, \{b\}, \{c\}$.

The join-irreducible elements that cover the bottom element are
called the \emph{atoms}. In Figure \ref{fig:posets}b, the bottom
element is $1$, the atoms are the prime numbers, and the
join-irreducible elements are powers of primes. Furthermore, two
primes $p$ and $q$ are relatively prime if their meet is the
bottom element $p \wedge q = \bot$, that is if $gcd(p,q) = 1$.
\footnote{This is reminiscent of the notation advocated by Graham,
Knuth, \& and Patashnik
\cite[p.115]{GrahamKnuthPatashnik:ConcreteMath} where $p \bot q$
denotes that $p$ and $q$ are relatively prime.} In Figure
\ref{fig:posets}d, the atoms are the the exhaustive set of
mutually exclusive assertions $a$, $k$ and $n$. The
join-irreducible elements are extremely important as all the
elements in the poset can be formed from joins of the
join-irreducible elements.

These examples show that the same set under two different ordering
relations can result in two different posets (Figures
\ref{fig:posets}a and b), and that two different sets under
different ordering relations can result in isomorphic posets
(Figures \ref{fig:posets}c and d). I have discussed these ideas
before \cite{Knuth:PhilTrans, Knuth:Questions}, and one can search
out more details in the accessible book by Davey \& Priestly
\cite{Davey&Priestley}, and the classic by Birkhoff
\cite{Birkhoff:1967}.

\subsection{Lattices}
Posets have the following properties. For a poset $P$, and
elements $a, b, c \in P$,
\begin{equation*}
\begin{array}{rll}
    P1. & For~~all~~a,~~a \leq a & \mbox{(\emph{Reflexive})}\\
    P2. & If~~a \leq b~~and~~b \leq a,~~then~~a = b &
    \mbox{(\emph{Antisymmetry})}\\
    P3. & If~~a \leq b~~and~~b \leq c,~~then~~a \leq c &
    \mbox{(\emph{Transitivity})}
\end{array}
\end{equation*}

If a unique meet $x \wedge y$ and join $x \vee y$ exists for all
$x, y \in P$, then the poset is called a \emph{lattice}. Each
lattice $L$ is actually an \emph{algebra} defined by the
operations $\vee$ and $\wedge$ along with any other relations
induced by the structure of the lattice. Dually, the operations of
the algebra uniquely determine the ordering relation, and hence
the lattice structure. Viewed as operations, the join and meet
obey the following properties for all $x, y, z \in L$
\begin{equation*}
\begin{array}{rll}
    L1. & x \vee x = x,~~~x \wedge x = x & \mbox{(\emph{Idempotency})}\\
    L2. & x \vee y = y \vee x,~~~x \wedge y = y \wedge x & \mbox{(\emph{Commutativity})}\\
    L3. & x \vee (y \vee z) = (x \vee y) \vee z,~~~x \wedge (y \wedge z) = (x \wedge y) \wedge z & \mbox{(\emph{Associativity})}\\
    L4. & x \vee (x \wedge y) = x \wedge (x \vee y) = x & \mbox{(\emph{Absorption})}\\
\end{array}
\end{equation*}

There is a special class of lattices called \emph{distributive
lattices} that follow
\begin{equation*}
\begin{array}{rll}
    D1. & x \wedge (y \vee z) = (x \wedge y) \vee (x \wedge z) & \mbox{(\emph{Distributivity of $\wedge$ over $\vee$})}\\
\end{array}
\end{equation*}
and its dual
\begin{equation*}
\begin{array}{rll}
    D2. & x \vee (y \wedge z) = (x \vee y) \wedge (x \vee z) & \mbox{(\emph{Distributivity of $\vee$ over $\wedge$})}\\
\end{array}
\end{equation*}
All distributive lattices can be expressed in terms of elements
consisting of sets where the join $\vee$ and the meet $\wedge$ are
identified as the set union $\cup$ and set intersection $\cap$,
respectively.

Some distributive lattices possess the property called
\emph{complementation} where for every element $x$ in the lattice,
there exists a unique element $\sim x$ such that
\begin{equation*}
\begin{array}{rl}
    C1. & x \vee \sim x = \top\\
    C2. & x \wedge \sim x = \bot\\
\end{array}
\end{equation*}
Boolean lattices are \emph{complemented distributive lattices}.
Since all distributive lattices can be described in terms of a
lattice of sets, the condition of a distributive lattice being
complemented is equivalent to the condition that the lattice
contains all possible subsets of the lattice elements, which is
called the \emph{powerset}. Thus lattices of powersets are Boolean
lattices. The situation gets interesting when one starts removing
elements from the powerset. For example, if one element is removed
from a Boolean lattice, then there will be another element that no
longer has a unique complement. This is equivalent to adding
constraints, and these constraints take the Boolean lattice to a
distributive lattice, which no longer has complementation as a
general property.

These are basic mathematical concepts and are not restricted to
the area of logical inference. Viewed as a collection of partially
ordered objects, we have a lattice. Viewed as a collection of
objects and a set of operations, such as $\vee$ and $\wedge$, we
have an algebra. What I will show in the remainder of this paper
is that a given lattice (or equivalently its algebra) can be
extended to a calculus using the methodology introduced by Cox,
and that there already exist a diverse array of examples outside
the realm of probability theory.

\section{Degrees of Inclusion}
For distinct $x$ and $y$ in a poset where $y$ includes $x$,
written $x \leq y$, it is clear that $x$ does not include $y$, $y
\nleq x$. However, we would like to generalize this notion of
inclusion so that even though $x$ does not include $y$, we can
describe the \emph{degree} to which $x$ includes $y$.  This idea
is perhaps made more clear by thinking about a concrete problem in
probability theory. Say that we know that a logical statement $a
\vee b \vee c$ is true. Clearly, $a \rightarrow a \vee b \vee c$
since $a \leq a \vee b \vee c$. However, it is useful to quantify
the \emph{degree} to which $a \vee b \vee c$ implies $a$, or
equivalently the degree to which $a$ includes $a \vee b \vee c$.
It is in this sense that we aim to generalize inclusion on a poset
to degrees of inclusion.

The goal of this paper is to emphasize that inclusion on a poset
can be generalized to degrees of inclusion, which results in a set
of rules or laws by which these degrees may be manipulated. These
rules are derived by requiring consistency with the structure of
the poset. At this stage, it is not clear for exactly what types
of posets or lattices this can be done, precisely what rules one
obtains and under what conditions, and exactly what types of
mathematical structures can be used to describe these degrees of
inclusion.  These remain open questions and will not be explicitly
considered in this paper.

What is clear is that Cox's basic methodology of deriving the sum
and product rules of probability from consistency requirements
with Boolean algebra \cite{Cox:1946, Cox:1961} has a much greater
range of applicability than he imagined. His specific results are
restricted to complemented lattices as he used the property of
complementation to derive the sum rule. In the following sections,
I will consider the larger class of distributive lattices, which
include Boolean lattices as a special case. To derive the laws
governing the manipulation of degrees of inclusion, I will rely on
the proofs introduced by Ariel Caticha \cite{Caticha:1998}, which
utilize consistency with associativity and distributivity.  As I
will show below, the sum rule is consistent with associativity of
the join, and therefore most likely enjoys a much greater range of
applicability---perhaps extending to all lattices.  The
derivations that follow will focus on finite lattices, although
extension to an infinite lattice is reasonable.

\subsection{Joins and the Sum Rule}
Throughout this subsection, I will follow and expand on Caticha's
derivation and maintain consistency with his basic notation, all
the while considering this endeavor as a generalization of
inclusion on a poset. We begin by defining a function $\phi$ that
assigns to a pair of elements $x$ and $y$ of the lattice $L$ a
real number\footnote{For simplicity I will work with degrees of
inclusion measured by real numbers. However, it should be kept in
mind that Caticha's derivations were developed for complex numbers
\cite{Caticha:1998}, Acz\'{e}l's solutions for the associativity
equation were for real numbers, groups and semigroups
\cite{Aczel:FunctEqns}, and Rota's theorems for the valuation
equation apply to commutative rings with an identity element
\cite{Rota:combinatorics}.}  $d \in \mathbb{R}$, so that $\phi:L
\times L \rightarrow \mathbb{R}$
\begin{equation}
d = \phi(x, y),
\end{equation}
and remember that $d$ represents the degree to which $x$ includes
$y$, which generalizes the algebraic inclusion $\leq$.
\footnote{This diverges from Caticha's development as he considers
functions that take a single poset element as its argument. I
discuss this difference in more detail in the sections that
follow.} This can be compared to Cox's notation where instead of
the function $\phi(x, y)$, he writes $(y \rightarrow x)$ for the
special case where he is considering implication among logical
statements. By replacing the comma with the solidus $\phi(x|y)$,
we obtain a notation more reminiscent of probability theory.

Now consider two join-irreducible elements of the lattice, $a$ and
$b$, where $a \wedge b = \bot$, and a third element $t$ such that
$a \leq t$ and $b \leq t$. We will consider the degree to which
the join $a \vee b$ of these join-irreducible elements includes
the element $t$. As $a \vee b$ is itself a lattice element, the
function $\phi$ allows us to describe the degree to which $a \vee
b$ includes $t$. This degree is written as $\phi(a \vee b, t)$. As
$a \wedge b = \bot$, this degree of inclusion can only be a
function of $\phi(a, t)$ and $\phi(b, t)$, which can be written as
\begin{equation}
\phi(a \vee b, t) = S(\phi(a, t),\phi(b, t)).
\end{equation}
Our goal is to determine the function $S$, which will tell us how
to use the degree to which $a$ includes $t$ and the degree to
which $b$ includes $t$ to compute the degree to which $a \vee b$
includes $t$. In this sense we are extending the algebra to a
calculus.

The function $S$ must maintain consistency with the lattice
structure, or equivalently with its associated algebra. If we now
consider another join-irreducible element $c \leq t$ where $a
\wedge c = \bot$ and $b \wedge c = \bot$, and form the lattice
element $(a \vee b) \vee c$, we can use associativity of the
lattice to write this element a second way
\begin{equation}
(a \vee b) \vee c = a \vee (b \vee c).
\end{equation}
Consistency with associativity requires that each expression gives
exactly the same result
\begin{equation}
S(\phi(a \vee b, t), \phi(c, t)) = S(\phi(a, t), \phi(b \vee c,
t)).
\end{equation}
Applying $S$ to the arguments $\phi(a \vee b, t)$ and $\phi(b \vee
c, t)$ above, we get
\begin{equation}
S(S(\phi(a, t),\phi(b, t)), \phi(c, t)) = S(\phi(a, t), S(\phi(b,
t),\phi(c, t))).
\end{equation}
This can be further simplified by letting $u = \phi(a, t)$, $v =
\phi(b, t)$, and $w = \phi(c, t)$, which gives
\begin{equation}
\label{eq:assocEqn}
S(S(u,v), w) = S(u, S(v,w)).
\end{equation}
This result is an equation for the function $S$, which emphasizes
its property of associativity. To people who are familiar with
Cox's work \cite{Cox:1946, Cox:1961}, this functional equation
should be immediately recognizable as what Acz\'{e}l appropriately
called \emph{the associativity equation}
\cite[pp.253-273]{Aczel:FunctEqns}. In Cox's derivation of
probability theory we are accustomed to seeing this in terms of
the logical conjunction. However, it is important to recognize
that both the conjunction and disjunction follow associativity,
and that this result generalized to the join is perfectly
reasonable. The general solution, from Acz\'{e}l
\cite{Aczel:FunctEqns}, is
\begin{equation}
S(u,v) = f(f^{-1}(u)+f^{-1}(v)),
\end{equation}
where $f$ is an arbitrary function. This can be simplified by
letting $g = f^{-1}$
\begin{equation}
g(S(u,v)) = g(u)+g(v),
\end{equation}
and writing this in terms of the original lattice elements we find
that
\begin{equation}
g(\phi(a \vee b, t)) = g(\phi(a, t)) + g(\phi(b, t)).
\end{equation}

As Caticha emphasizes, this result is remarkable, as it reveals
that there exists a function $g:\mathbb{R} \rightarrow \mathbb{R}$
re-mapping these numbers to a more convenient representation. Thus
we can define a new map from the lattice elements to the real
numbers, such that $\nu(a, t) \equiv g(\phi(a, t))$. This lets us
write the combination rule as
\begin{equation}
\nu(a \vee b, t) = \nu(a, t) + \nu(b, t),
\end{equation}
which is the familiar \emph{sum rule} of probability theory for
mutually exclusive (join-irreducible) assertions $a$ and $b$
\begin{equation}
p(a \vee b | t) = p(a | t) + p(b | t).
\end{equation}

This result is extremely important, as I have made no reference at
all to probability theory in the derivation. Only consistency with
associativity of the join, which is a property of all lattices,
has been used. This means that for a given lattice, we can define
a mapping from a pair of its elements to a real number, and when
we take joins of its join-irreducible elements, we can compute the
new value of this join by taking the sum of the two numbers. There
are some interesting restrictions, which I will discuss later.

\subsubsection{Extending the Sum Rule}
The assumption made above was that the two lattice elements were
join-irreducible and that their meet was the bottom element. How
do we perform the computation in the event that this is not the
case? In this section I will demonstrate how the sum rule can be
extended in a distributive lattice. Consider two lattice elements
$x$ and $y$. In a distributive lattice all elements can be written
as a unique join of join-irreducible elements
\begin{equation}
x = (\bigvee_{i=1}^{n} p_i) \vee (\bigvee_{i=1}^{k} q_i)
\end{equation}
and
\begin{equation}
y = (\bigvee_{i=1}^{m} r_i) \vee (\bigvee_{i=1}^{k} q_i),
\end{equation}
where I have written them so that the join-irreducible elements
they have in common are the elements $q_1, q_2, ... q_k$. The join
of $x$ and $y$ can be written as
\begin{equation}
x \vee y = (\bigvee_{i=1}^{n} p_i) \vee (\bigvee_{i=1}^{k} q_i)
\vee (\bigvee_{i=1}^{m} r_i) \vee (\bigvee_{i=1}^{k} q_i),
\end{equation}
which can be simplified to
\begin{equation}
x \vee y = (\bigvee_{i=1}^{n} p_i) \vee (\bigvee_{i=1}^{k} q_i)
\vee (\bigvee_{i=1}^{m} r_i),
\end{equation}
since by $L1$ and $L2$
\begin{equation}
(\bigvee_{i=1}^{k} q_i) \vee (\bigvee_{i=1}^{k} q_i) =
\bigvee_{i=1}^{k} q_i.
\end{equation}
Since the ${p_i}, {q_i},$ and ${r_i}$ are all join-irreducible
elements, we can use the sum rule repeatedly to obtain
\begin{equation}
\nu(x \vee y, t) = {\sum_{i = 1}^{n} \nu(p_i, t)} + {\sum_{i =
1}^{k} \nu(q_i, t)} + {\sum_{i = 1}^{m} \nu(r_i, t)}.
\end{equation}
Notice that the first two terms on the right are $\nu(x, t)$. I
will add two more terms to the right (which cancel one another),
and then group the terms conveniently
\begin{equation}
\nu(x \vee y, t) = ({\sum_{i = 1}^{n} \nu(p_i, t)} + {\sum_{i =
1}^{k} \nu(q_i, t)}) + ({\sum_{i = 1}^{m} \nu(r_i, t)} + {\sum_{i
= 1}^{k} \nu(q_i, t)}) - {\sum_{i = 1}^{k} \nu(q_i, t)}.
\end{equation}
This can be further simplified to give
\begin{equation}
\nu(x \vee y, t) = \nu(x, t) + \nu(y, t) - {\sum_{i = 1}^{k}
\nu(q_i, t)},
\end{equation}
where we have the original sum rule minus a cross-term of sorts,
which avoids double-counting the join-irreducible elements. The
lattice elements forming this additional term can be found from
\begin{equation}
x \wedge y = \bigvee_{i=1}^{k} q_i,
\end{equation}
so that
\begin{equation}
\nu(x \wedge y, t) = {\sum_{i = 1}^{k} \nu(q_i, t)}.
\end{equation}
Note that to maintain consistency with the original sum rule, we
must require that
\begin{equation}
\nu(\bot, t) = 0.
\end{equation}
This allows us to write the generalized sum rule as
\begin{equation}
\nu(x \vee y, t) = \nu(x, t) + \nu(y, t) - \nu(x \wedge y, t).
\label{eq:gensum}
\end{equation}
What is remarkable is that we have derived a result that is valid
for all distributive lattices! When joins of more than two
elements are considered, this procedure can be iterated to avoid
double-counting the join-irreducible elements that the elements
share. Changing notation slightly this equation is identical to
the general sum rule for probability theory.
\begin{equation}
p(x \vee y | t) = p(x | t) + p(y | t) - p(x \wedge y | t).
\label{eq:sum}
\end{equation}

\subsubsection{Valuations}
I recently found that these ideas have been developing
independently in geometry and combinatorics with influence of
Gian-Carlo Rota \cite{Rota:GeoProb, Rota:combinatorics,
Klain&Rota}. A \emph{valuation} is \textit{defined} on a lattice
of sets (distributive lattice) as a function $v: L \rightarrow
\mathbb{A}$ that takes a lattice element to an element of a
commutative ring with identity $\mathbb{A}$, and satisfies
\begin{equation}
v(a \vee b) + v(a \wedge b) = v(a) + v(b),
\end{equation}
where $a,b \in L$, and
\begin{equation}
v(\bot) = 0.
\end{equation}
By subtracting $v(a \wedge b)$ from both sides, we see that the
valuation equation is just the generalized sum rule
(Eqn.~\ref{eq:gensum})
\begin{equation}
v(a \vee b) = v(a) + v(b) - v(a \wedge b).
\end{equation}
When applied to Boolean lattices, valuations are called measures.
As far as I am aware, the valuation equation was defined by
mathematicians and not derived. However, following Cox and
Caticha, we have derived it directly from the sum rule, which was
derived from consistency with associativity of the join.

Note that valuations, as we describe them here have a single
argument and do not explicitly consider the degree to which one
lattice element includes a second lattice element. This is not a
problem, as one can define bi-valuations, tri-valuations, and
multi-valuations although it is not clear how to interpret all of
these functions as generalizations of inclusion on a poset. One
can define valuations that represent the degree to which an
element $x$ includes $\top$, by
\begin{equation}
v(x) \equiv \nu(x,\top),
\end{equation}
which can be interpreted in probability theory as a prior
probability \footnote{In this notation $\top$ refers to the
truism, which is the join of all possible statements. In the past,
I have preferred to write probabilities in the style of Jaynes
where $I$ is used to represent our `prior information' as in
$p(x|I)$. The truism represents this prior information in part,
since we know \textit{a priori} that one of the exhaustive set of
mutually exclusive assertions is true. However, excluded in the
notation $p(x|\top)$ is explicit reference to the part of the
prior information $I$ that is relevant to the probability
assignments. I choose to leave $I$ out of the probability notation
here to emphasize the fact that the function $p$ takes two lattice
elements as arguments---not abstract symbols like $I$.}
\begin{equation}
v(x) \equiv p(x|\top).
\end{equation}
However, throughout this paper I will work with bi-valuations, as
they can be used to represent the degree to which one lattice
element includes another.

\subsubsection{M\"{o}bius Functions}
I have demonstrated how the sum rule can be extended for
distributive lattices, but how is this handled in general for
posets where associativity holds? One must rely on what is called
the M\"{o}bius function for the poset. I will begin by discussing
a special class of real-valued functions of two variables defined
on a poset, such as $f(x,y)$, which are non-zero only when $x \leq
y$. This set of functions comprises the \emph{incidence algebra}
of the poset \cite{Rota:foundations}. The sum of two functions $h
= f + g$ is defined the usual way by
\begin{equation}
h(x,y) = f(x,y) + g(x,y),
\end{equation}
as is multiplication by a scalar $h = \lambda f$. However, the
product of two functions in the incidence algebra is found by
taking the convolution over the interval of elements in the poset
\begin{equation}
h(x,y) = \sum_{x \leq z \leq y}{f(x,z) g(z,y)}.
\end{equation}

We can define three useful functions \cite{Rota:foundations,
Krishnamurthy:combinatorics}
\begin{eqnarray}
\label{eq:delta} \delta(x,y) & = &
   \left\{ \begin{array}{rl}
       1 & if~~x = y \\
       0 & if~~x \neq y \end{array} \right.
   \mbox{(\emph{Kronecker delta function})} \\
\nonumber \\
\label{eq:incidence} n(x,y) & = &
   \left\{ \begin{array}{rl}
       1 & if~~x < y \\
       0 & if~~x \nless y \end{array} \right.
   \mbox{(\emph{incidence function})} \\
\nonumber \\
\label{eq:zeta} \zeta(x,y) & = &
   \left\{ \begin{array}{rl}
       1 & if~~x \leq y \\
       0 & if~~x \nleq y \end{array} \right.
   \mbox{(\emph{zeta function})}
\end{eqnarray}
The delta function indicates when two poset elements are equal.
The incidence function indicates when an element $x$ is properly
contained in an element $y$. Last, the zeta function indicates
whether $y$ includes $x$, which means that the zeta function is
equal to the sum of the delta function and the incidence function
\begin{equation}
\zeta(x,y) = n(x,y) + \delta(x,y).
\end{equation}

It is important to be able to invert functions in the incidence
algebra. For example, the inverse of the zeta function, $\mu(x,y)$
satisfies
\begin{equation}
\label{eq:zetaInversion} \sum_{x \leq z \leq y}{\zeta(x,z)
\mu(z,y)} = \delta(x,y).
\end{equation}
One can show \cite{Rota:combinatorics, Rota:foundations,
Barnabei&Pezzoli} that the function $\mu(x,y)$, called the
\emph{M\"{o}bius function}, is defined by
\begin{eqnarray}
\label{eq:mobius}
& \mu(x,x) = 1 & x \in P \nonumber\\
& \sum_{x \leq z \leq y}{\mu(x,z)} = 0 & x < y \\
& \mu(x,y) = 0 & if~~x \nleq y, \nonumber
\end{eqnarray}
where
\begin{equation}
\label{eq:mobiusCondition} \sum_{x \leq z \leq y}{\mu(x,z)} =
\sum_{x \leq z \leq y}{\mu(z,y)}.
\end{equation}
Rather than providing a proof, I will demonstrate this by
considering the possible cases. Obviously, if $x = y$ then
\begin{equation}
\zeta(y,y) \mu(y,y) = 1,
\end{equation}
which is consistent with the first condition for the M\"{o}bius
function (\ref{eq:mobius}). Next, if $x \leq y$ we can use the
fact that $\zeta(x,z) = 1$ only when $x \leq z$ to rewrite
(\ref{eq:zetaInversion})
\begin{equation}
\sum_{x \leq z \leq y}{\zeta(x,z) \mu(z,y)} = \delta(x,y).
\end{equation}
as
\begin{equation}
\sum_{x \leq z \leq y}{\mu(z,y)} = \delta(x,y).
\end{equation}
The sum can be rearranged using (\ref{eq:mobiusCondition})
\begin{equation}
\sum_{x \leq z \leq y}{\mu(x,z)} = \delta(x,y),
\end{equation}
which is consistent with the second condition (\ref{eq:mobius})
when $x < y$. Last, if $x > y$ then (\ref{eq:zetaInversion}) is
trivially satisfied.

More importantly, the M\"{o}bius function is used to invert
valuations on a poset $P$ \cite{Rota:combinatorics,
Rota:foundations, Barnabei&Pezzoli} so that given a function
\begin{equation}
\label{eq:gx}
g(x) = \sum_{d \leq x}{f(d)}
\end{equation}
one can find $f(x)$ by
\begin{equation}
\label{eq:fx}
f(x) = \sum_{d \leq x}{\mu(d,x) g(d)}.
\end{equation}
What's going on here is made more clear \cite{Barnabei&Pezzoli} by
considering the poset of natural numbers $N$ in the usual order
(see Fig \ref{fig:posets}a). This poset is totally ordered and the
M\"{o}bius function is easily found to be
\begin{equation}
\label{eq:muN} \mu_N(x,y) =
   \left\{ \begin{array}{rl}
       1 & if~~x = y \\
       -1 & if~~x \prec y \\
       0 & otherwise \end{array} \right.
\end{equation}
Given
\begin{equation}
\label{eq:sums}
g(n) = \sum_{m \leq n}{f(m)}
\end{equation}
we find, using (\ref{eq:fx}) and (\ref{eq:muN}), that
\begin{equation}
\label{eq:differences}
f(n) = g(n) - g(n-1).
\end{equation}
This is the finite difference operator, which is the discrete
analog of the \emph{fundamental theorem of calculus} for a poset
\cite{Rota:foundations, Barnabei&Pezzoli}, which basically relates
sums (\ref{eq:sums}) to differences (\ref{eq:differences}).

Readers well-versed in number theory have seen the classic
M\"{o}bius function \cite{Mobius1832} used to invert the Riemann
zeta function
\begin{equation}
\zeta(s) = \sum_{n=1}^{\infty}{\frac{1}{n^s}},
\end{equation}
which is important in the study of prime numbers (refer to the
poset in Fig.~\ref{fig:posets}b). The M\"{o}bius function in this
case is defined
\cite{Rota:foundations,Barnabei&Pezzoli,GrahamKnuthPatashnik:ConcreteMath}
as
\begin{equation}
\begin{array}{rll}
& \mu(1) = 1 \\
& \sum_{d|m}{\mu(d)} = 0
\end{array}
\end{equation}
where the sum is over all numbers $d$ dividing $m$ , that is all
numbers $d \leq m$ in the poset in Fig.~\ref{fig:posets}b.
Specifically, its values are found by
\begin{equation}
\label{eq:muPRIMES} \mu(d) =
   \left\{ \begin{array}{rl}
       0 & if~~d~~is~~divisible~~by~~some~~p^2\\
       (-1)^k & if~~is~~a~~product~~of~~k~~distinct~~primes \end{array} \right.
\end{equation}
where $p$ is a prime. This leads to the inverse of the zeta
function given by
\begin{equation}
\frac{1}{\zeta(s)} = \sum_{n=1}^{\infty}{\frac{\mu(n)}{n^s}}.
\end{equation}

Clearly, order theory and the incidence algebra ties together
areas of mathematics such as the calculus of finite differences
and number theory. In the next section, I will show that we can
use the M\"{o}bius function to extend the sum rule over the entire
poset.

\subsubsection{The Inclusion-Exclusion Principle}
By iterating the generalized sum rule, we obtain what Rota calls
the \emph{inclusion-exclusion principle} \cite[p.7]{Klain&Rota}
\begin{equation}
\nu(x_1 \vee x_2 \vee \cdots \vee x_n, t) = \sum_i{\nu(x_i,t)} -
\sum_{i<j}{\nu(x_i \wedge x_j, t)} + \sum_{i<j<k}{\nu(x_i \wedge
x_j \wedge x_k, t)} - \cdots \label{eq:incl-excl}
\end{equation}
This equation holds for distributive lattices where every lattice
element can be written as a unique join of elements. As I will
show, this principle appears over and over again, and is a sign
that order-theoretic principles and distributivity underlie the
laws having this form.

Rota showed that the inclusion-exclusion principle can be obtained
from the M\"{o}bius function of the poset \cite{Rota:foundations,
Barnabei&Pezzoli}. For a Boolean lattice of sets the M\"{o}bius
function is given by
\begin{equation}
\label{eq:muBOOLE} \mu_{B}(x,y) = (-1)^{|y|-|x|}
\end{equation}
whenever $x \subseteq y$ and $0$ otherwise, where $|x|$ is the
cardinality of the set $x$. The M\"{o}bius function for a
distributive lattice is similar
\begin{equation}
\label{eq:muDIST} \mu_{D}(x,y) =
   \left\{ \begin{array}{rl}
       1 & if~~x = y \\
       (-1)^n & if~~x < y \\
       0 & if~~x \nleq y \end{array} \right.
\end{equation}
where in the second case $y$ is the join of $n$ distinct elements
covering $x$. This leads directly to the alternating sum and
difference in the inclusion-exclusion principle as one sums down
the lattice.

The inclusion-exclusion principle appears in a delightful variety
of contexts, several of which will be explored later. We have
already seen it in the context of the sum rule of probability
theory
\begin{equation}
p(x \vee y | t) = p(x | t) + p(y | t) - p(x \wedge y | t).
\end{equation}
It also appears in Cox's generalized entropies \cite{Cox:1961},
which were explored earlier in McGill's multivariate
generalization of mutual information \cite{McGill:1955}, and
examined more recently as co-information by Tony Bell
\cite{Bell:2003}. The familiar mutual information makes the basic
point
\begin{equation}
I(x,y) = H(x) + H(y) - H(x, y).
\end{equation}
Again, this is the inclusion-exclusion principle at work. Rota
\cite{Rota:combinatorics} gives an interesting example from
P\"{o}lya and Szeg\"{o} \cite[Vol II., p. 121]{Polya&Szego} which
I will shorten here to
\begin{equation}
max(a, b) = a + b - min(a, b),
\end{equation}
where $a$ and $b$ are real numbers. This equation may seem
horribly obvious, but like the others, it can be extended by
iterating. The idea is to include and exclude down the lattice!
The inclusion-exclusion pattern is an important clue that order
theory plays an important role.

\subsection{Assigning Valuations}
There are some interesting and useful results on assigning
valuations and extending them to larger lattices (eg.
\cite[p.9]{Klain&Rota}). The fact that probabilities are
valuations implies that these results are relevant to assigning
probabilities. Specifically, Rota \cite[Theorem 1, Corollary 2,
p.35]{Rota:combinatorics} showed that

\begin{quotation}
A valuation in a finite distributive lattice is uniquely
determined by the values it takes on the set of join-irreducibles
of $L$, and these values can be arbitrarily assigned.
\end{quotation}

By considering $p(x|\top)$, this result can be applied to the
assignment of prior probabilities. In the lattice of logical
assertions ordered by logical implication, the join-irreducible
elements are the exhaustive set of mutually exclusive assertions.
Thus by assigning their prior probabilities, the probabilities of
their various disjunctions are uniquely determined. This is easy,
just use the sum rule.

More profound is the fact that Rota's theorem states that these
assignments can be \emph{arbitrary}. This means that there is no
information in the Boolean algebra of these assertions, and hence
the inferential calculus, to guide us in these assignments. Thus
probability theory tells us nothing about assigning priors. Other
principles, such as symmetry, constraints, and consistency with
other aspects of the problem \emph{must} be employed to assign
priors. Once the priors are assigned, order-theoretic principles
dictate the remaining probabilities through the inferential
calculus.

Last, it is not clear to me how to assign valuations in posets
where at least one element can be written as the join of
join-irreducible elements in more than one way, such as in the
lattice $\mathbf{N}_5$ \cite[Fig. 4.3(i)]{Davey&Priestley}. Once
again, consistency must be the guiding principle. When there is
more than one way to write an element as a join of
join-irreducible elements, the valuations assigned to those
elements must be consistent with the particular sum rule for that
lattice.  This is not an issue for probability theory, but it will
become an issue when this methodology is extended to other
problems.

\subsection{Meets and the Product Rule}
The product rule is usually seen as being necessary for computing
probabilities of conjunctions of logical statements, whereas the
sum rule is necessary for computing the probabilities of
disjunctions. This actually isn't true. The sum rule allows one to
compute the probabilities of conjunctions equally well.
Rearranging Equation~\ref{eq:sum} gives
\begin{equation}
p(x \wedge y | t) = p(x | t) + p(y | t) - p(x \vee y | t),
\end{equation}
or more generally
\begin{equation}
\nu(x \wedge y, t) = \nu(x, t) + \nu(y, t) - \nu(x \vee y, t).
\end{equation}
The important point is that for some problems, this just isn't
useful.

The key to understanding the product rule is to realize that there
are actually two kinds of logical conjunctions in probability
theory. The first is the `\emph{and}' that is implemented by the
meet on the Boolean lattice. In the earlier example on who stole
the tarts, this type of conjunction leads to statements like $(k
\vee a) \wedge (n \vee a)$, which can be read literally as
`\textit{Somebody stole the tarts \textbf{and} it wasn't the
Knave!}' The meet performs the role of the logical conjunction
while working \emph{within} the hypothesis space. The second type
of logical conjunction occurs when one \emph{combines} two
hypothesis spaces to make a bigger space. For example, we can
combine a lattice $F$ describing different types of fruit, with a
lattice $Q$ describing the quality of the fruit by taking a
Cartesian product of $F \times Q$, which results in statements
like `\textit{The fruit is an apple \textbf{and} it is spoiled!}'
This type of conjunction is not readily computed using the sum
rule. To handle both types of logical conjunctions, we will derive
the product rule. In the section on quantum mechanics, we will see
that these ideas are not limited to probability theory.

\begin{figure}[htbp]
\label{fig:prod} \caption{a. The product of two Boolean lattices
$F$ and $Q$ can be constructed by taking the Cartesian product of
their respective elements. b. This leads to the product lattice $P
= F \times Q$, which represents both spaces jointly. However, $P$
contains seven statements (gray circles), which belong to an
equivalence class of absurd statements. c. By grouping these
elements in the equivalence class under $\bot$, we can form a new
lattice $\tilde{P}$, which is a subdirect product of $F$ and $Q$.
d. An alternate lattice structure can be formed by treating the
statements represented by the join-irreducible elements of the
subdirect product $\tilde{P}$ as atomic statements and forming
their powerset. This yields a Boolean lattice $B$ distinct from,
yet isomorphic to the Cartesian product $P$. Both logical
structures $\tilde{P}$ and $B$ are implicitly used in probability
theory, and both follow the distributive law $D1$.}
\includegraphics[height=.8\textheight]{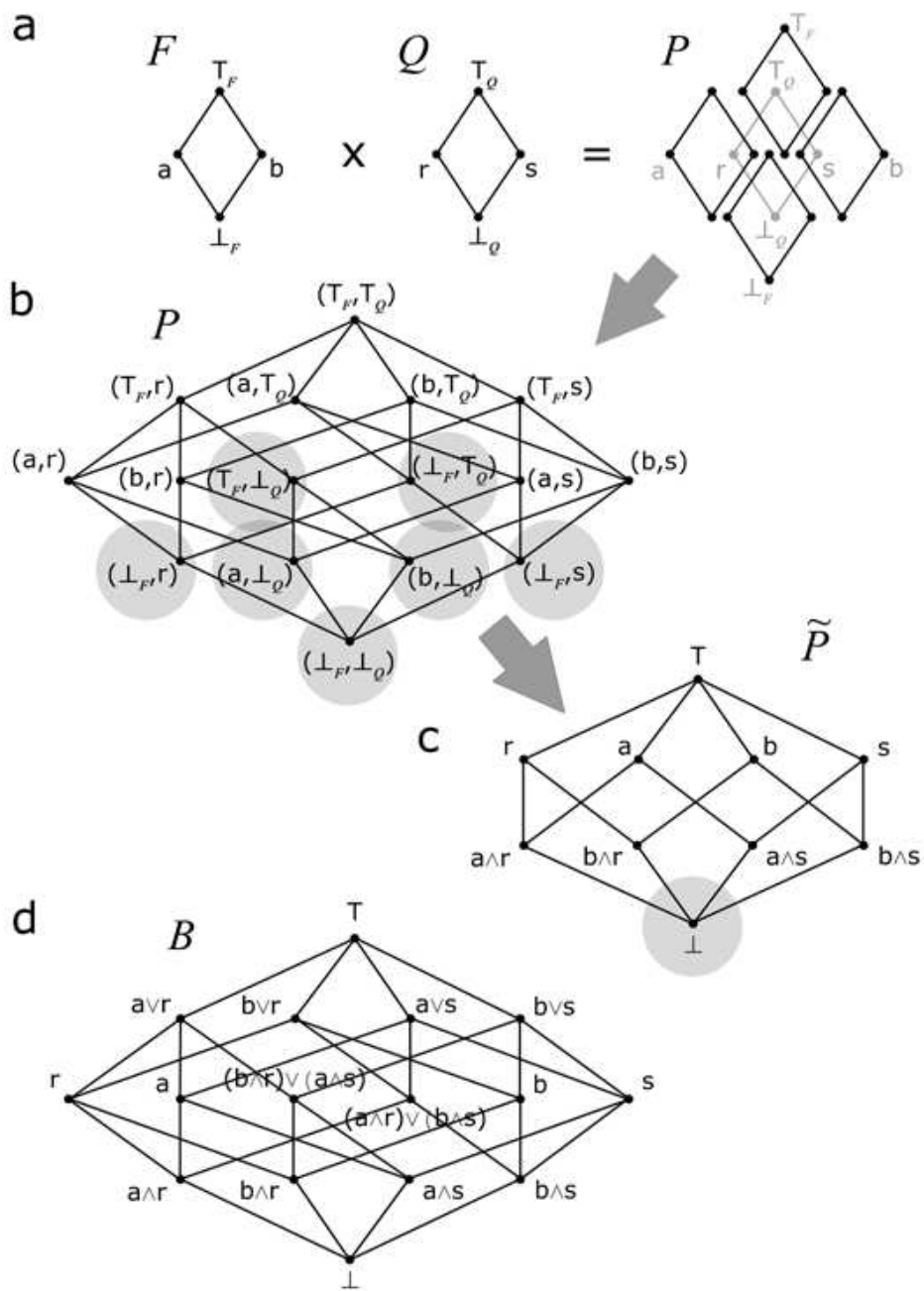}
\end{figure}

\subsubsection{The Lattice Product}
We can define the lattice $F$ describing the type of fruit by
specifying the two atomic assertions covering the bottom
\begin{align*}
a &= \textit{`It is an apple!'}\\
b &= \textit{`It is a banana!'}.
\end{align*}
This will form a Boolean lattice with four elements shown on the
left-side of Figure~\ref{fig:prod}a. This Boolean lattice
structure with two atoms is denoted by $\mathbf{2}^2$. As usual,
the top element stands for the truism, which is a logical
statement that says `\textit{It is an apple or a banana!}', which
we write symbolically as $\top_F = a \vee b$. The bottom is the
absurdity which says `\textit{It is an apple and a banana!}',
written $\bot_F = a \wedge b$. The lattice is clearly Boolean as
the complement of $a$ is $b$, and vice versa (i.e. $a \wedge b =
\bot_F$ and $a \vee b = \top_F$).

Similarly, what is known about the quality of the fruit can be
described by the lattice $Q$ generated by the atomic assertions
\begin{align*}
r &= \textit{`It is ripe!'}\\
s &= \textit{`It is spoiled!'}.
\end{align*}
These assertions generate the center lattice in
Figure~\ref{fig:prod}a, which is also Boolean (isomorphic to
$\mathbf{2}^2$).

We can combine these two lattices by taking the \emph{lattice
product}. Graphically, this can be constructed by fixing one of
the two lattices, and placing copies of the other over each
element of the former. In Figure \ref{fig:prod}a, I fix the
lattice $Q$ and place copies of the lattice $F$ over each element
of $Q$. The elements of the product lattice $P = F \times Q$ are
found by forming the Cartesian product of the elements of the two
lattices. For example, the element $(a,r)$ represents the logical
statement that says `\textit{The fruit is an apple and it is
ripe!}' Two elements of $F \times Q$, $(f_1, q_1)$ and $(f_2,
q_2)$, can be ordered coordinate-wise
\cite[p.42]{Davey&Priestley}, so that
\begin{equation}
(f_1, q_1) \leq (f_2, q_2)~~~~\mbox{when}~~~f_1 \leq
f_2~~~\mbox{and}~~~q_1 \leq q_2,
\end{equation}
which leads to a coordinate-wise definition of $\vee$ and $\wedge$
\begin{equation}
(f_1, q_1) \vee (f_2, q_2) = (f_1 \vee f_2, q_1 \vee q_2)
\end{equation}
\begin{equation}
(f_1, q_1) \wedge (f_2, q_2) = (f_1 \wedge f_2, q_1 \wedge q_2).
\end{equation}
The result is the Boolean lattice $P$ in Figure~\ref{fig:prod}b,
which is isomorphic to $\mathbf{2}^2 \times \mathbf{2}^2 =
\mathbf{2}^4$. I should make the meaning of some of these
statements more explicit. For example, $(\top_F,r)$ is a statement
that says `\textit{It is an apple or a banana and it is ripe!}'
Similarly, $(b,\top_Q)$ is a statement that says `\textit{It is a
banana that is either ripe or spoiled!}'

The lattice product is associative, so that for three lattices
$J$, $K$, and $L$
\begin{equation}
J \times (K \times L) = (J \times K) \times L.
\end{equation}
This associativity translates to the associativity of the meet of
the Cartesian products of the elements, which is consistent with
the fact that the lattice product of two lattices is a lattice.

What is interesting is that there are seven elements (gray
circles) that involve at least one of the two absurdities. These
seven elements, $(\top_F,\bot_Q)$, $(\bot_F,\top_Q)$,
$(\bot_F,r)$, $(a,\bot_Q)$, $(b,\bot_Q)$, $(\bot_F,s)$, and
$(\bot_F,\bot_F)$ each belong to an \emph{equivalence class} of
logically absurd statements as they say things like `\textit{It is
ripe and spoiled!}' and `\textit{It is an apple and a banana!}' If
we group these absurd statements together under a new symbol
$\bot$, we can construct the lattice $\tilde{P}$ in
Figure~\ref{fig:prod}c. I have simplified the element labels so
that the element $b$ stands for $(b,\top_Q)$, which can be read as
`\textit{The fruit is a banana!}', and $s$ stands for
$(\top_F,s)$, which can be read as `\textit{The fruit is
spoiled!}' The lattice $\tilde{P}$ is a \emph{subdirect product}
of the lattices $F$ and $Q$ since $\tilde{P}$ can be embedded into
their Cartesian product $P$, and its projections onto both $F$ and
$Q$ are surjective (onto, but not one-to-one).

The symbol $\wedge$ again represents the meet in the subdirect
product $\tilde{P}$, so that the meet of $a$ and $s$ gives the
element $a \wedge s$, which is equivalent to a spoiled apple
$(a,s)$. In this way we see that the meet in the subdirect product
lattice behaves like the meet in the original hypothesis space,
while simultaneously implementing the Cartesian product. However,
there are some key differences. First, $\tilde{P}$ is not Boolean.
For example, there is no unique complement to the statement $a
\wedge s$, as both $b$ and $r$ satisfy the requirements for the
complement
\begin{equation*}
\begin{array}{r}
(a \wedge s) \vee b = \top \\
(a \wedge s) \wedge b = \bot \\
\end{array}
\end{equation*}
and
\begin{equation*}
\begin{array}{r}
(a \wedge s) \vee r = \top \\
(a \wedge s) \wedge r = \bot \\
\end{array}
\end{equation*}

A more important difference is the fact that the meet in $P$
follows both distributive laws $D1$ and $D2$, whereas the meet in
$\tilde{P}$ follows only $D1$. To demonstrate this consider the
following in the context of $\tilde{P}$
\begin{equation}
a \wedge (r \vee s) = a \wedge \top = a,
\end{equation}
and
\begin{equation}
(a \wedge r) \vee (a \wedge s) = a,
\end{equation}
which is consistent with $D1$
\begin{equation}
a \wedge (r \vee s) = (a \wedge r) \vee (a \wedge s).
\end{equation}
Now consider
\begin{equation}
a \vee (r \wedge s) = a \vee \bot = a,
\end{equation}
whereas
\begin{equation}
(a \vee r) \wedge (a \vee s) = \top \wedge \top = \top,
\end{equation}
which is inconsistent with $D2$, which is distributivity of $\vee$
over $\wedge$. The difficulty here is clear. In the Cartesian
product $P$ and the subdirect product $\tilde{P}$, statements like
$a \vee r$ and $a \vee s$ do not have distinct interpretations
since one cannot use the Cartesian product to form the statement
`\textit{It is an apple or the fruit is ripe!}' Another way to
look at the loss of property $D2$ is to notice that because we
have identified the bottom elements in the equivalence relation we
lose distributivity of $\vee$ over $\wedge$ (property $D2$), and
maintain distributivity of $\wedge$ over $\vee$ (property $D1$).
Had we identified the top elements we would have obtained the dual
situation.

There is one last important construction we can perform.  We can
construct another lattice by considering the join-irreducible
statements of the subdirect product differently. If we let $a
\wedge r$, $a \wedge s$, $b \wedge r$, $b \wedge s$ represent an
atomic set of exhaustive mutually exclusive statements rather than
a Cartesian product of statements, then all other elements in the
lattice can be formed from the joins of these atoms. The result is
a Boolean lattice $B$ that is isomorphic to the lattice product
$P$, so that $B \sim P \sim \mathbf{2}^4$. In this lattice the
atomic statements, such as $a \wedge r$, do not represent
Cartesian products, but instead represent elementary statements
like `\textit{It is a ripe apple!}'. For this reason, $B$ contains
logical statements not included in the Cartesian product $P$ or
the subdirect product $\tilde{P}$. For example, we can construct
statements like $(b \wedge r) \vee (a \wedge s)$, which state
`\textit{It is either a ripe banana or a spoiled apple!}' Lattices
formed this way naturally follow both $D1$ and $D2$, as they are
Boolean.

The important point here is that we use each of these constructs
in different applications of probability theory without explicit
consideration. Most relevant is the fact that each of these three
lattice constructs $P$, $\tilde{P}$, and $B$ follows $D1$. Thus if
we require that our calculus satisfies distributivity of $\wedge$
over $\vee$ then we will have a rule that is consistent with
properties of each of the logical constructs we have considered
here.

\subsubsection{Deriving the Product Rule}
The product rule is important as it gives us a way to compute the
degree of inclusion for meets of elements in a lattice constructed
from the product of two distributive lattices. We look for a
function $P$ that allows us to write the degree to which the meet
of two elements $x \wedge y$ includes a third element $t$ without
relying on the join $x \vee y$. Cox chose the form
\begin{equation}
\nu(x \wedge y, t) = P(\nu(x, t), \nu(y, x \wedge t)),
\label{eq:prodform}
\end{equation}
which was later shown by Tribus \cite{Tribus:Rational} and Smith
\& Erickson \cite{Smith&Erickson} to be the only functional form
that satisfies consistency with associativity of $\wedge$.
\footnote{I use the sum rule in the following derivation, which
requires that I use the mapping $\nu$ rather than $\phi$.}

Consider the elements $a$ and $b$ where $a \wedge b = \bot$, and
the elements $r$ and $s$ where $r \wedge s = \bot$. We reproduce
Caticha's derivation \cite{Caticha:1998} and consider
distributivity $D1$ of the meet over the join in the lattice
product
\begin{equation}
(a, (r \vee s)) \equiv a \wedge (r \vee s) = (a \wedge r) \vee (a
\wedge s).
\end{equation}
This equation gives us two different ways to express the same
poset element. Consistency with distributivity $D1$ requires that
the same value is associated with each of these two expressions.
Using the sum rule (\ref{eq:sum}) and the form of $P$ consistent
with associativity (\ref{eq:prodform}), we find that
distributivity requires that
\begin{equation}
P(\nu(a, t), \nu(r \vee s, a \wedge t)) = \nu(a \wedge r, t) +
\nu(a \wedge s, t),
\end{equation}
which further simplifies to
\begin{equation}
P(\nu(a, t), \nu(r, a \wedge t) + \nu(s, a \wedge t)) = P(\nu(a,
t), \nu(r, a \wedge t)) + P(\nu(a, t), \nu(s, a \wedge t)).
\end{equation}
If we let $u = \nu(a,t)$, $v = \nu(r, a \wedge t)$, and $w =
\nu(s, a \wedge t)$, the equation above can be written more
concisely as
\begin{equation}
P(u, v + w) = P(u, v) + P(u, w).
\label{eq:prod-d1-1}
\end{equation}
This is a functional equation for the function $P$, which captures
the essence of distributivity. By working with this equation, we
will obtain the functional form of $P$.

The idea is to show that $P(u, v + w)$ is linear in its second
argument.  If we let $z = w + v$, and write (\ref{eq:prod-d1-1})
as
\begin{equation}
P(u, z) = P(u, v) + P(u, w),
\end{equation}
we can look at the second derivative with respect to $z$. Using
the chain rule we find that
\begin{equation}
{{\partial} \over {\partial v}} = {{\partial z} \over {\partial
v}} {{\partial} \over {\partial z}} = {{\partial} \over {\partial
z}}.
\end{equation}
This can be done also for $w$ giving
\begin{equation}
{{\partial} \over {\partial v}} = {{\partial} \over {\partial w}}
= {{\partial} \over {\partial z}}.
\end{equation}
Writing the second derivative with respect to $z$ as
\begin{equation}
{{\partial}^2 \over {\partial z^2}} = {{\partial} \over {\partial
v}} {{\partial} \over {\partial w}},
\end{equation}
we find that
\begin{equation}
\begin{array}{rll}
{{\partial^2} \over {\partial z^2}} P(u,z) & = & {{\partial} \over
{\partial v}} {{\partial} \over {\partial w}} (P(u, v) + P(u, w))\\
& = & {{\partial} \over {\partial v}} ({{\partial} \over {\partial w}} P(u, w))\\
& = & {{\partial} \over {\partial w}} ({{\partial} \over {\partial v}} P(u, w))\\
& = & 0,
\end{array}
\end{equation}
which means that the function $P$ is linear in its second argument
\begin{equation}
P(u,v) = A(u)v + B(u).
\label{eq:linear-first}
\end{equation}
If (\ref{eq:linear-first}) is substituted back into
(\ref{eq:prod-d1-1}) we find that $B(u) = 0$.

We can use $D1$ another way by considering $(a \vee b) \wedge r$.
This leads to a condition that looks like
\begin{equation}
P(v + w, u) = P(v, u) + P(w, u),
\label{eq:prod-d1-2}
\end{equation}
where $u$, $v$, $w$ are appropriately redefined. Following the
approach above, we find that $P$ is also linear in its first
argument
\begin{equation}
P(u,v) = A(v)u. \label{eq:linear-second}
\end{equation}
Together with (\ref{eq:linear-first}), the general solution is
\begin{equation}
P(u,v) = Cuv,
\end{equation}
where $C$ is an arbitrary constant. Thus, we have the
\emph{product rule}
\begin{equation}
\nu(x \wedge y, t) = C \nu(x, t) \nu(y, x \wedge t),
\label{eq:product-rule}
\end{equation}
which tells us the degree to which the new element $x \wedge y$
includes the element $t$. This looks more familiar if we set $C =
1$ and re-write the rule in probability-theoretic notation,
\begin{equation}
p(x \wedge y | t) = p(x | t) p(y | x \wedge t).
\end{equation}

If the lattice we are working in is formed by the lattice product,
(\ref{eq:product-rule}) can be rewritten using the Cartesian
product notation
\begin{equation}
\nu((x, y), (t_x, t_y)) = C \nu((x, \top_y), (t_x, t_y))
\nu((\top_x, y), (x \wedge t_x, t_y)),
\end{equation}
where $x \wedge y \equiv (x, y)$, $t \equiv (t_x, t_y)$, $x \equiv
(x,\top_y)$, and $y \equiv (\top_x,y)$. Simplifying, we see that
\begin{equation}
\nu((x, y), (t_x, t_y)) = C \nu(x, t_x) \nu(y, t_y),
\label{eq:product-rule2}
\end{equation}
where $\nu(x, t_x)$ is computed in one lattice and $\nu(y, t_y)$
in the other.

\section{Bayes' Theorem}
The origin of Bayes' Theorem is perhaps the most well-known. It
derives directly from the product rule and consistency with
commutativity of the meet. When it is true that $x \wedge y$ = $y
\wedge x$, one can write the product rule two ways
\begin{equation*}
\nu(x \wedge y, t) = C \nu(x, t) \nu(y, x \wedge t)
\end{equation*}
and
\begin{equation*}
\nu(y \wedge x, t) = C \nu(y, t) \nu(x, y \wedge t).
\end{equation*}
Consistency with commutativity requires that these two results are
equal. Setting them equal and rearranging the terms leads to
\emph{Bayes' Theorem}
\begin{equation}
\nu(x, y \wedge t) = {{\nu(x, t) \nu(y, x \wedge t)} \over {\nu(y,
t)}}.
\end{equation}
This will look more familiar if we let $x \equiv h$ be a logical
statement representing a hypothesis, and let $y \equiv d$ be a new
piece of data, and change notation to that used in probability
theory
\begin{equation}
p(h | d \wedge t) = {{p(h | t) p(d | h \wedge t)} \over {p(d |
t)}}.
\end{equation}
This makes more sense now that it is clear that the two statements
$h$ and $d$ come from different lattices. This is why $d$
represents a \textit{new} piece of data, it represents information
we were not privy to when we implicitly constructed the lattice
including the hypothesis $h$. The statements $h$ and $d$ belong to
two different logical structures and Bayes' Theorem tells us how
to do the computation when we combine them.
    To perform this computation, we need to first make assignments.
The prior assignment $p(h | t)$ is made in the original lattice of
hypotheses, whereas the likelihood assignment $p(d | h \wedge t)$
is made in the product lattice. The evidence, while usually not
assigned, refers to an assignment that would take place in the
original data lattice.

\section{Laws from Order}
Why is all this important? Because, the sum rules are associated
with all lattices, and sum and product rules are not just
associated with Boolean algebra, but with distributive algebras.
This is a much wider range of application than was ever considered
by Cox, as the following examples demonstrate.

\subsection{Information Theory from Order}
Most relevant to Cox is the further development of the calculus of
inquiry \cite{Cox:1979, Fry:CourseNotes, Fry:Cybernetics,
Knuth:Questions, Knuth:PhilTrans}, which appears to be a natural
generalization of information theory. The extension of Cox's
methodology to distributive lattices in general is extremely
important to this development, as the lattice structure of
questions is not a Boolean lattice, but is instead the \emph{free
distributive lattice} \cite{Knuth:Questions, Knuth:PhilTrans}.
This free distributive lattice of questions is generated from the
ordered set of down-sets of its corresponding Boolean lattice of
assertions. A probability measure on the Boolean assertion lattice
induces a valuation, which we call \emph{relevance} or
\emph{bearing}, on the question lattice. I will show in a future
paper \cite{Knuth:inquiry} that the join-irreducible questions,
called \emph{elementary questions} by Fry \cite{Fry:CourseNotes},
have relevances that are equal to $-p \log p$, called the
\emph{surprise}, where $p$ is the probability of the assertion
defining the elementary question. Joins of these elementary
questions form real questions, which via the sum rule have
valuations formed from sums of surprises, or \emph{entropy}. Going
further up the question lattice, one uses the generalized sum
rule, which results in mutual information, and eventually
generalized entropy \cite{Cox:1961}, also called co-information
\cite{Bell:2003}. The lattice structure is given by the algebra of
questions, and the generalized entropies are the valuations
induced on the question lattice by the probability measure
assigned to the Boolean algebra of assertions. Exactly how these
valuations are induced, manipulated, and used in calculations
regarding questions will be discussed elsewhere
\cite{Knuth:inquiry}.

\subsection{Geometry from Order}
Perhaps more interesting is the fact that much of geometry can be
derived from these order-theoretic concepts. There has been much
work done in this area, which is called \emph{geometric
probability}. This invariably calls up thoughts of Buffon's Needle
problem, but the range of applications is much greater. I have
found the introductory book by Klain and Rota \cite{Klain&Rota} to
be very useful, and below I discuss several illustrative examples
from their text.

\subsubsection{The Lattice of Parallelotopes}
Imagine a Cartesian coordinate system in \textit{n}-dimensions. We
will consider orthogonal parallelotopes, which are rectangular
boxes with sides parallel to the coordinate axes. By taking finite
unions (joins) and intersections (meets) of these parallelotopes,
we can construct the lattice (or equivalently the algebra) of
\textit{n}-dimensional parallelotopes $Par(n)$
\cite[p.30]{Klain&Rota}. In one dimension, the parallelotope is
simply a closed interval on the real line.

To extend the algebra to a calculus, to each parallelotope $P$ we
will assign a valuation\footnote{Note that one can consider
$\mu(P) \equiv \mu(P,T)$, where $T$ refers to a parallelotope to
which the measure is in some sense normalized.} $\mu(P)$, which is
invariant with respect to geometric symmetries relevant to
$Par(n)$, such as invariance with respect to translations along
the coordinate system and permutations of the coordinate labels.
Assigning valuations that are consistent with these important
geometric symmetries is analogous to Jaynes' use of group
invariance to derive prior probabilities \cite{Jaynes:Prior}.

\begin{figure}
\label{fig:geometry} \caption{An illustration of the join and meet
of two orthogonal parallelotopes.}
\includegraphics[height=.4\textheight]{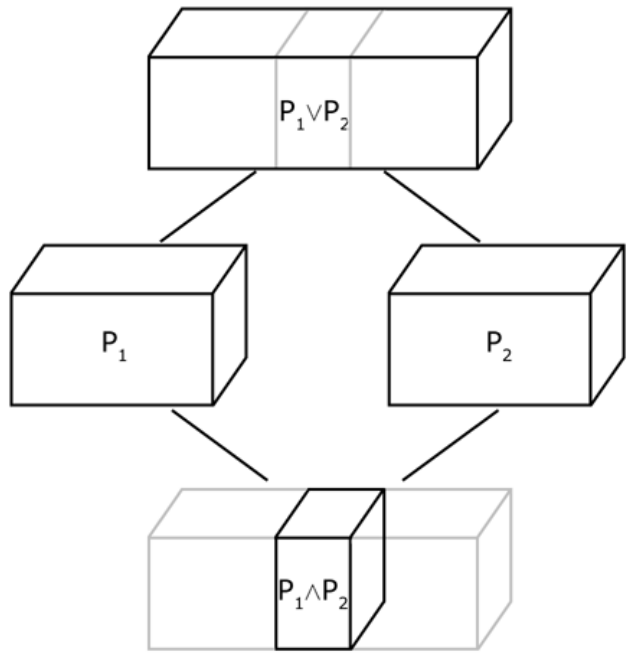}
\end{figure}

Before looking at these invariant valuations in more detail,
Figure \ref{fig:geometry} shows the join of two parallelotopes
$P_1$ and $P_2$. Since this lattice is distributive, the sum rule
can be used to compute the valuation of the join $P_1 \vee P_2$
\begin{equation}
\mu(P_1 \vee P_2) = \mu(P_1) + \mu(P_2) - \mu(P_1 \wedge P_2).
\end{equation}
Again we see the familiar inclusion-exclusion principle.

\subsubsection{A Basis Set of Invariant Valuations}
At this point, you have probably already identified a valuation
that will satisfy invariance with respect to translation and
coordinate label permutation. One obvious valuation for
three-dimensional parallelotopes suggested by the illustration is
\emph{volume}
\begin{equation}
\mu_3(x) = x_1 x_2 x_3,
\end{equation}
where $x_1$, $x_2$ and $x_3$ are the side-lengths of the
parallelotope. Surprisingly, this is not the only valuation that
satisfies the invariance properties we are considering. There is
also a valuation, which is proportional to the \emph{surface area}
\begin{equation}
\mu_2(x) = x_1 x_2 + x_2 x_3 + x_3 x_1,
\end{equation}
which is easily shown to satisfy both the invariance properties as
well as the sum rule. In fact, there is a basis set of invariant
valuations, which in the case of three-dimensional parallelotopes
consists of the volume, surface area, \emph{mean width}
\begin{equation}
\mu_1(x) = x_1 + x_2 + x_3,
\end{equation}
and the \emph{Euler characteristic} $\mu_0$, which for
parallelotopes is equal to one for non-empty parallelotopes and
zero otherwise. The fact that these valuations form a basis means
that we can write any valuation as a linear combination of these
basis valuations
\begin{equation}
\mu = a \mu_3 + b \mu_2 + c \mu_1 + d \mu_0.
\end{equation}
In general, it is not clear under which invariance conditions one
obtains a basis set of valuations rather than a unique functional
form. This is an extremely important issue when we consider the
issue of assigning prior probabilities in probability theory.
Jaynes' demonstrated how to derive priors that are consistent with
certain invariances, and cautioned that if the number of
parameters in the transformation group is less than the number of
model parameters, then the prior will only be partially determined
\cite{Jaynes:Prior}. How to consistently assign priors in this
case is an open problem.

Furthermore, the Euler characteristic is interesting in that it
takes on discrete rather than continuous values. This is something
that is not seen in measure theory indicating that this
development is more general than the typical measure-theoretic
approaches. An important example of this has been identified
within the context of information theory. Acz\'{e}l
\cite{aczel+etal:natural} showed that the Hartley entropy
\cite{hartley}, which takes on only discrete values, has certain
`\emph{natural}' properties shared only with the Shannon entropy
\cite{Shannon&Weaver}. This will also be discussed in more detail
elsewhere \cite{Knuth:inquiry}.

\subsubsection{The Euler Characteristic}
The Euler characteristic appears in other lattices and is perhaps
best known from the lattice of convex polyhedra where it satisfies
the following formula
\begin{equation}
\mu_0 = F - E + V,
\end{equation}
where $F$ is the number of faces, $E$ is the number of edges, and
$V$ is the number of vertices \cite{Rota:GeoProb}. For all convex
polyhedra in three-dimensions, $\mu_0 = 2$. For example, if we
consider a cube, we see that it has 6 faces, 12 edges, and 8
vertices, so that $6 - 12 + 8 = 2$. Again, this is an example of
the inclusion-exclusion principle, which comes from the sum rule.
In the lattice of simplicial complexes, a face is a 2-simplex, an
edge is a 1-simplex, and a vertex is a 0-simplex. To compute the
characteristic, we add at one level, subtract at the next, and add
at the next and so on. This geometric law derives directly from
order theory via the sum rule.

\subsubsection{Spherical Triangles}
The connections with order theory do not stop with polyhedra, but
extend into continuous geometry. Klain \& Rota
\cite[p.158]{Klain&Rota} show that the solid angle subtended by a
triangle inscribed on a sphere, called the \emph{spherical
excess}, can be found using the inclusion-exclusion principle
\begin{equation}
\Omega(\Delta) = \alpha + \beta + \gamma - \pi
\end{equation}
where $\alpha, \beta, \gamma \in [0, \pi]$ denote the angles of
the spherical triangle. Such examples highlight the degree to
which order theory dictates laws.

\subsection{Statistical Physics from Order}
Up to this point, probability theory and geometry have been the
main examples by which I have demonstrated the use of
order-theoretic principles to derive a calculus from an algebra.
Thanks to the efforts of Ed Jaynes \cite{Jaynes:InfoTheory}, Myron
Tribus \cite{Tribus:Thermodynamics} and others, I am able to wave
my hands and state that statistical physics derives from
order-theoretic principles. In one important respect this argument
is a sham, and that is where entropy is concerned. The
\emph{principle of maximum entropy} \cite{Jaynes:InfoTheory,
Jaynes:MaxEnt}, which is central to statistical physics, lies just
beyond the scope of this order-theoretic framework. It is possible
that a fully-developed calculus of inquiry \cite{Knuth:inquiry}
will provide useful insights. With entropies being associated with
the question lattice, application of the principle of maximum
entropy to enforce consistency with known constraints may in some
sense be dual to the \textit{maximum a posteriori} procedure in
probability theory. However, at this stage it is certain that the
probability-based features of the theory of statistical physics
derive directly from these order-theoretic principles.

\subsection{Quantum Mechanics from Order}
Most surprising is Ariel Caticha's realization that the rules for
manipulating quantum mechanical amplitudes in slit experiments
derive from consistency with associativity and distributivity of
experimental setups \cite{Caticha:1998}.  Each experimental setup
describes an idealized experiment that describes a particle
prepared at an initial point and detected at a final point. These
experimental setups are simplified so that the only observable
considered is that of position. The design of each setup
constrains the statements that one can make about the particle.

These setups can be combined in two ways, which I will show are
essentially \emph{meets} and \emph{joins} of elements of a poset.
However, there are additional constraints on these operations
imposed by the physics of the problem. We will see that the meets
are not commutative, and this makes these algebraic operations
significantly different from the AND and OR of propositional
logic. This lack of commutativity means that there is no Bayes'
Theorem analog for quantum mechanical amplitudes. Furthermore, the
operation of negation is never introduced---nor is it necessary.
This sets Caticha's approach apart from other quantum logic
approaches where the negation of a quantum mechanical proposition
remains a necessary concept.

Caticha considers a simple case where the only experiments that
can be performed are those that detect the local presence or
absence of a particle. He considers a particle, which is prepared
at time $t_i$ at position $x_i$ and is later detected at time
$t_f$ at position $x_f$. Experimental setups of this sort will
test statements like `\textit{the particle moved from $x_i$ to
$x_f$}'. The physics becomes interesting when we place obstacles
in the path of the particle. For example, we can place a barrier
with a single slit at position $x_1$. The detector at $x_f$ will
only detect the particle if it moves from $x_i$ to $x_1$ and then
onward to $x_f$, where $t_i < t_1 < t_f$. \footnote{These paths
are in space-time. For simplicity, I will often omit reference to
the time coordinate.} This barrier with a slit imposes a
constraint on the particles that can be detected. The central idea
is that experimental setups can be combined in two ways:
increasing the number of constraints on the particle behavior, or
by decreasing the number of constraints. This allows one to impose
an ordering relation on the experimental setups, and by
considering the set of all possible setups where the particle is
prepared at $x_i$ and detected at $x_f$ we have a poset of
experimental setups.

\begin{figure}
\label{fig:setups} \caption{A partial schematic of Caticha's poset
of experimental setups where a particle is prepared at $x_i$ and
is later detected at $x_f$. Solid lines indicate $\prec$, whereas
dashed lines indicate $\leq$ where there are setups that are not
illustrated. See the text for details.}
\includegraphics{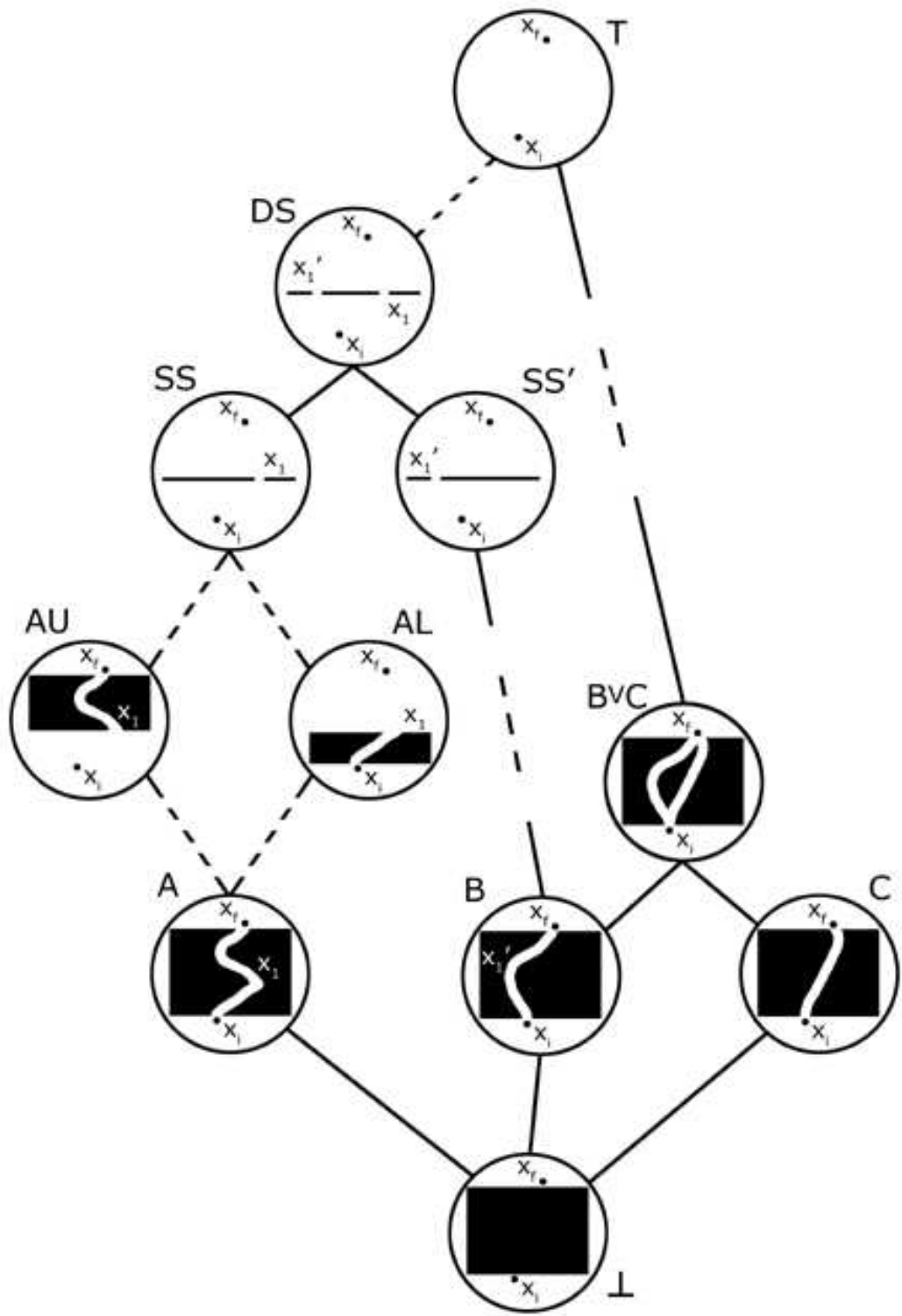}
\end{figure}

Setups with fewer constraints give greater freedom to the possible
behavior of the detected particle. Of course, the ordering
relation we choose can be defined in terms of the constraints on
the particle's behavior or, dually, on the freedom that the setup
allows the particle. Each relation will lead to a poset, which is
dual to the poset ordered by its dual ordering relation.
Maintaining a positive attitude, I will use the ordering relation
that describes the freedom that the setup allows, so that setups
with greater freedom will include setups that allow only a portion
of this freedom. Thus at the top of this poset
(Figure~\ref{fig:setups}) is the setup $\top$ that has no
obstructions placed between $x_i$ and $x_f$. The particle having
been prepared at $x_i$ and detected at $x_f$ is free to have taken
any path during its traversal. Caticha uses a concise notation to
describe setups. The top setup can be written as $\top \equiv
[x_f, x_i]$, which reading right to left states that the particle
is known to start at $x_i$ and is known to arrive at $x_f$.

The most constrained setup is the one at the bottom, $\bot$ , as
it is completely filled with obstructions.\footnote{All setups
with obstructions that provide no way for a particle to travel
from $x_i$ to $x_f$ belong to the same equivalence class of
obstructed setups and are represented by $\bot$. This includes
setups with obstructed pathways.} This setup is analogous to the
logical absurdity in the sense that there is no way for a particle
prepared at $x_i$ to be detected at $x_f$. More interesting are
the join-irreducible setups that cover $\bot$. These are the
setups that allow only a single path from $x_i$ to $x_f$. Three
examples $A, B,$ and $C$ can be seen in Figure~\ref{fig:setups}.

All the setups in the poset can be generated by joining the
join-irreducible setups. As Caticha, defined his poset in terms of
an algebra, we must look at this algebra carefully to identify the
complete set of join-irreducible setups. Obviously the atomic
setups, which are the setups with a single path from $x_i$ to
$x_f$, are join-irreducible. However, Caticha defined the join of
two setups only for cases where the two setups differ by at most
one point at one time. Thus we must work through his algebra using
the poset structure to discover how to perform the join in more
general cases. Consider the two setups $SS$ and $SS'$ in
Figure~\ref{fig:setups}. The setup $SS$ is a single-slit
experiment where the particle is known to have been prepared at
$x_i$, is known to have gone through the slit at $x_1$ and was
detected at $x_f$, written concisely as $SS = [x_f, x_1, x_i]$.
Similarly, $SS'$ is a different single-slit setup written $SS' =
[x_f, x_1', x_i]$. Their join is a double-slit setup
\begin{equation}
DS = SS \vee SS'
\end{equation}
found by
\begin{equation}
DS = [x_f, x_1, x_i] \vee [x_f, x_1', x_i],
\end{equation}
which is written concisely as
\begin{equation}
DS = [x_f, (x_1, x_1'), x_i].
\end{equation}
This can be read as: `\textit{The particle was prepared at $x_i$,
is known to have passed through either the slit at $x_1$ or the
slit at $x_1'$, and was detected at $x_f$}'. This definition
describes the join of two setups differing at only one point.
Before generalizing the join, we will first examine the meet.

Caticha describes the simplest instance of a meet \cite[Eqn.
2]{Caticha:1998}, which is
\begin{equation}
\label{eq:meet}
[x_f, x_1] \wedge [x_1, x_i] = [x_f, x_1, x_i],
\end{equation}
giving the single-slit experiment $SS$ in Figure~\ref{fig:setups}.
This equation is interesting, because neither setup on the
left-hand side of (\ref{eq:meet}) is a valid setup in our poset,
since the particle is not always prepared at $x_i$ and it is not
always detected at $x_f$. Instead, what is going on here is that
this meet represents the Cartesian product of two setups. Two
smaller setup posets are being combined to create a larger setup
poset. This is demonstrated by the meet of AU and AL in
Figure~\ref{fig:setups}, each of which has a single path on one
side of $x_1$ and is free of obstructions on the other side. AU is
the Cartesian product of the top setup in the set of posets
describing particles prepared at $x_i$ and detected at $x_1$,
which I write as $\top_{1i}$, and a setup consisting of a single
path in the set of posets describing particles prepared at $x_1$
and detected at $x_f$. We can write these posets coordinate-wise
\begin{equation}
\begin{array}{rll}
AL & \equiv & (\top_{f1}, A_{1i}) \\
AU & \equiv & (A_{f1}, \top_{1i})
\end{array}
\end{equation}
where $A_{1i}$ is the setup where the particle is prepared at
$x_i$ and is detected at $x_1$ and is constrained to follow that
portion of the path in the single-path setup $A$. Setups $A_{f1}$
and $\top_{f1}$ are defined similarly. Their meet is found
trivially, since for each coordinate, $x \wedge \top = x$ for all
$x$. Thus
\begin{equation}
AU \wedge AL = (A_{f1}, A_{1i})
\end{equation}
which is a valid expression for the entire single-path setup $A$
\begin{equation}
A = (A_{f1}, A_{1i}),
\end{equation}
and is consistent with Caticha's definition in (\ref{eq:meet}).
This interpretation, which should be compared to our earlier
discussion on the lattice product, also provides insight into how
one can decompose the single-path setups $A$, $B$, and $C$ into a
series of infinitesimal displacements. Each infinitesimal
displacement of a particle can be described using a setup in a
smaller poset. Using the Cartesian product, these setups can be
combined to form larger and larger displacements. Thus the
single-path setups can be seen to represent meets of many shorter
single-path segment setups.

We can use what we have learned from the product space example to
better understand the join of two experimental setups. We can join
the setups $AL$ and $AU$ using the Cartesian product notation.
Since $\top$ is the top element of the poset, when joined with any
other element of the poset the result is always the top, i.e.
$\top \vee x = \top$, for any $x$ in the poset. Thus we see that
\begin{equation}
\begin{array}{rll}
AU \vee AL & = & (A_{f1}, \top_{1i}) \vee (\top_{f1}, A_{1i}) \\
& = & (A_{f1} \vee \top_{f1}, \top_{1i} \vee A_{1i}) \\
& = & (\top_{f1}, \top_{1i}) \\
& \equiv & SS.
\end{array}
\end{equation}
The interpretation of the two posets forming this Cartesian
product is key to better understanding the join. The result states
that the particle is prepared at $x_i$ and is free to travel
however it likes to be detected at $x_1$, and that it is prepared
at $x_1$ and is free to travel however it likes resulting in a
detection at $x_f$. The result is a particle that is prepared at
$x_i$, passes through $x_1$ and is detected at $x_f$. Thus their
join is the single slit experiment, $AU \vee AL = SS$. This result
extends Caticha's algebraic definition of the join.

Two setups with non-intersecting paths can also be joined.  This
must be the case since the top setup in this poset is
obstacle-free and includes all single-path setups. Consider a
setup divided by the straight line connecting $x_i$ and $x_f$. Two
setups can be formed from this, one by filling the left half with
an obstacle, call it $L$, and the other by filling the right half
$R$. Clearly, their join must be the top element, as one setup
prevents the particle from being to the left of the line and the
other one prevents it from being to the right. This can be
extended by considering two setups each with paths that do not
intersect one another. This example is shown in
Figure~\ref{fig:setups}, as the join of two single-path setups $B$
and $C$, which results in a two-path setup $B \vee C$.

At this stage, it is not completely clear to me how to handle
setups with two paths that intersect at multiple points. Clearly,
two paths that intersect at a single median point, such as $x_1$,
can be considered to be the product of two setups each with two
non-intersecting paths. Again, consistency will be the guiding
principle. What is important to this present development is that
we know enough of the algebra to see what kind of laws we can
derive from order theory by generalizing from order-theoretic
inclusion to degrees of inclusion. First, Caticha showed that the
join is associative. This implies that there exists a sum rule.
Second, he showed that when the setups exist the meet is
associative and distributes over the join. This leads to a product
rule. By measuring degrees of inclusion among experimental setups
with complex numbers, Caticha showed that the sum and product
rules applied to complex-valued valuations are consistent with the
rules used to manipulate quantum amplitudes. By looking at the
join of setups $B$ and $C$ one can visualize how application of
the sum rule leads to Feynman path integrals \cite{Feynman:1948},
which can be used to compute amplitudes for setups like $B \vee
C$, and by iterating across the poset, $SS$ and $\top$. Also, the
amplitude corresponding to a setup representing any finite
displacement can be decomposed into a product of amplitudes for
setups representing smaller successive displacements. Last, it
should be noted that the lack of commutativity of the meet implies
that there is no Bayes' Theorem analog for quantum amplitudes.

Furthermore, using these rules Caticha showed that that the only
consistent way to describe time evolution is by a linear
Schr\"{o}dinger equation. This is a truly remarkable result, as no
information about the physics of the particle was used in this
derivation of the setup calculus---only order theory! The physics
comes in when one assigns values to the setups in the poset. This
is done by assigning values to the infinitesimal setups, which is
equivalent to assigning the priors in probability theory. At the
stage of assigning amplitudes, we can now only rely on symmetry,
constraints, and consistency with other aspects of the problem.
The calculus of setup amplitudes will handle the rest. The fact
that these assignments rely on the Hamiltonian and that they are
complex-valued are now the key issues. Looking more closely at the
particular symmetries, constraints and consistencies that result
in the traditional assignments independent of the setup calculus
will provide deeper insight into quantum mechanics and will teach
us much about how to extend this new methodology into other areas
of science and mathematics.

\section{Discussion}
Probability theory is often summarized by the statement
`\emph{probability is a sigma algebra}', which is a concise
description of the mathematical properties of probability.
However, descriptions like this can limit the way in which one
thinks about probability in much the same way that the statement
`\emph{gravity is a vector field}' limits the way one thinks about
gravitation. To gain new understanding in an area of study, the
foundations need to be re-explored. Richard Cox's investigations
into the role of consistency of probability theory with Boolean
algebra were a crucial step in this new exploration. While Cox's
technique has been celebrated in several circles within the area
of probability theory \cite{Jaynes:InfoTheory, Tribus:Rational,
Jaynes:Book}, the deeper connections to order theory discussed
here have not yet been fully recognized. The exception is the area
of geometric probability where Gian-Carlo Rota has championed the
importance of valuations on posets giving rise to an area of
mathematics which ties together number theory, combinatorics, and
geometry. Simple relations, such as the inclusion-exclusion
principle, act as beacons signalling that order theory lies not
far beneath.

Order theory dictates the way in which we can extend an algebra to
a calculus by assigning numerical values to pairs of elements of a
poset to describe the degree to which one element includes
another. Consistency here is the guiding principle. The sum rule
derives directly from consistency with associativity of the join
operation in the algebra. Whereas, the product rule derives from
consistency with associativity of the meet, and consistency with
distributivity of the meet over the join.

It is clear that the basic methodology of extending an algebra to
a calculus, which is presently explicitly utilized in probability
theory and geometric probability, is implicitly utilized in
information theory, statistical mechanics, and quantum mechanics.
The order-theoretic foundations suggest that this methodology
might be used to extend any class of problems where partial
orderings (or rankings) can be imposed to a full-blown calculus.
One new example explored at this meeting is the \emph{ranking of
preferences} in decision theory, which is explored in Ali Abbas'
contribution to this volume \cite{Abbas:Preferences}. More obvious
is the relevance of this methodology to the development of the
calculus of inquiry \cite{Cox:1979, Fry:CourseNotes,
Knuth:Questions, Knuth:PhilTrans}, as well as Bob Fry's extension
of this calculus to cybernetic control \cite{Fry:Cybernetics}. A
serious study of the relationship between order theory and
geometric algebra, recognized and noted by David Hestenes
\cite{Hestenes+Ziegler, Hestenes:LinAlg+Geom}, is certain to yield
important new results. With the aid of geometric algebra, an
examination of projective geometry in this order-theoretic context
may provide new insights into Carlos Rodr\'{i}guez's observation
that the cross-ratio of projective geometry acts like Bayes'
Theorem \cite{Carlos:1991}.

\begin{theacknowledgments}
This work supported by the NASA IDU/IS/CICT Program and the NASA
Aerospace Technology Enterprise. I am deeply indebted to Ariel
Caticha, Bob Fry, Carlos Rodr\'{i}guez, Janos Acz\'{e}l, Ray
Smith, Myron Tribus, David Hestenes, Larry Bretthorst, Jeffrey
Jewell, Domhnull Granquist-Fraser, and Bernd Fischer for
insightful and inspiring discussions, and many invaluable remarks
and comments.
\end{theacknowledgments}


\doingARLO[\bibliographystyle{aipproc}]
          {\ifthenelse{\equal{\AIPcitestyleselect}{num}}
             {\bibliographystyle{arlonum}}
             {\bibliographystyle{arlobib}}
          }
\bibliography{knuth}

\end{document}